\documentclass[aps,pra,reprint,groupedaddress,amssymb,amsmath]{revtex4-1}

\newcommand{\CC}{\mathbb{C}}

\newcommand{\ket}[1]{\left| #1 \right>} 
 
\newcommand{\vspan}[1]{\mathrm{span}\left\{ #1 \right\}}

\begin{document}


\title{The inductive entanglement classification yields ten rather than eight classes of four-qubit entangled states}


\author{Miriam Backens}
\email[]{m.backens@bristol.ac.uk}
\affiliation{School of Mathematics, University of Bristol, UK}



\begin{abstract}
 Lamata et al.\ use an inductive approach to classify the entangled pure states of four qubits under stochastic local operations and classical communication (SLOCC) \cite{lamata_inductive_2007}. The inductive method yields a priori ten different entanglement superclasses, of which they discard three as empty. One of the remaining superclasses is split in two, resulting in eight superclasses of genuine four-qubit entanglement.
 
 Here, we show that two of the three discarded superclasses are in fact non-empty and should have been retained. We give explicit expressions for the canonical states for those superclasses, up to SLOCC and qubit permutations. Furthermore, we confirm that the third discarded superclass is indeed empty, yielding a total of ten superclasses of genuine four-qubit entanglement under the inductive classification scheme.
\end{abstract}


\maketitle

\section{Introduction}

Entangled states are an important resource in quantum information and computation. The classification of entanglement under stochastic local operations and classical communication (SLOCC) aims to group quantum states according to which quantum information tasks they can accomplish \cite{bennett_exact_2000}. This classification is an important question in quantum theory, and it has not been resolved for more than four qubits.

This is in part because there are infinitely many distinct entanglement classes under SLOCC for four or more qubits \cite{dur_three_2000}. Approaches to the classification of four-qubit entanglement group these infinitely many classes into `superclasses', which contain entanglement classes with similar structure that are nevertheless distinct under SLOCC \cite{lamata_inductive_2007,verstraete_four_2002}. In the following, we sometimes use the word `class' to refer to a superclass, it should be clear  from context which is meant.

One scheme for entanglement classification on multi-qubit states is the inductive approach by Lamata et al.\ \cite{lamata_inductive_2006}. In a subsequent paper from 2007, the same authors apply this approach to the classification of four-qubit states \cite{lamata_inductive_2007}, reporting eight distinct entanglement superclasses of genuine four-partite entanglement (up to qubit permutations). In the process, they consider three other potential superclasses, discarding them with the claim that they are empty. We point out that in two cases this is erroneous: the classes called $\mathfrak{W}_{0_k\Psi,W}$ and $\mathfrak{W}_{W,W}$ in the naming scheme from \cite{lamata_inductive_2007} should not have been discarded because they do contain states that do not fall into any other entanglement superclass. For these two entanglement superclasses, we show not only that they are non-empty but also derive their respective canonical states, into which any state in the given class can be transformed by SLOCC. This is analogous to the approach taken for the other entanglement classes in \cite{lamata_inductive_2007}. We furthermore confirm that in the third case -- that of the class called $\mathfrak{W}_{GHZ,GHZ}$ -- Lamata et al.\ were correct in discarding it.

Throughout this paper, we use the equational rules for the classification of three-qubit entanglement derived by Li et al.\ in 2006 \cite{li_simple_2006}.

In the following, we first recap the inductive entanglement classification in Section \ref{s:inductive_classification} and the equational rules in Section \ref{s:equational_classification}. The new results are contained in Section \ref{s:results}.

\section{The inductive entanglement classification\label{s:inductive_classification}}

In the inductive entanglement classification by Lamata et al., introduced in \cite{lamata_inductive_2006} and applied to four-qubit states in \cite{lamata_inductive_2007}, entanglement classes are distinguished as follows: Consider an $n$-qubit state $\ket{\psi}$. Let $\ket{v_0}$ and $\ket{v_1}$ be two linearly independent states of the first qubit. Then the full state can be written as:
\begin{equation}\label{eq:decomposition}
 \ket{\psi} = \ket{v_0}\ket{\phi_0}+\ket{v_1}\ket{\phi_1}
\end{equation}
for some $(n-1)$-qubit states $\ket{\phi_0},\ket{\phi_1}$. Note that $\ket{v_0}$, $\ket{v_1}$ do not need to be normalised, and even if they are chosen to be normalised, $\ket{\phi_0}$ and $\ket{\phi_1}$ need not be. As SLOCC operations can change the norm of states, we work with unnormalised states in general anyway.

The entanglement class -- or rather superclass, as there are infinitely many different SLOCC classes on four or more qubits -- is determined by the types of $(n-1)$-qubit entangled vectors found in different spanning sets for $\vspan{\ket{\phi_0},\ket{\phi_1}}$. Those in turn are given by the entanglement classes of $(n-1)$-qubit states: hence the inductiveness of the classification process.

By convention, states are classified according to the spanning sets for  $\vspan{\ket{\phi_0},\ket{\phi_1}}$ containing vectors with the `least amount of entanglement'.
In the case of the classification of four-qubit entanglement, the chosen order on the entanglement classes of three-qubit states is as follows \cite{lamata_inductive_2007}:
\begin{equation}
 000 < 0\Psi < GHZ, W,
\end{equation}
where furthermore GHZ is usually considered before W.
Here, `$000$' denotes a fully separable state, `$0\Psi$' denotes a state which is the product of a single-qubit state and an entangled two-qubit state. This type of state is sometimes referred to as a `bipartite separable state'. GHZ and W are the usual classes of fully entangled three-qubit states \cite{dur_three_2000}, the (unnormalised) standard representatives of which we write as $\ket{GHZ}=\ket{000}+\ket{111}$ and $\ket{W}=\ket{001}+\ket{010}+\ket{100}$, respectively.

Lamata et al.\ label the entanglement classes according to the types of entangled vectors in the spanning set as $\mathfrak{W}_{X,Y}$ where $X,Y$ take values `$000$', `$0\Psi$', `$GHZ$' or `$W$'. In the case of $0\Psi$, the bipartition may be specified by a subscript, i.e.\ $0_k\Psi$, where $k\in\{1,2,3\}$, denotes a state where the $k$-th qubit is in a product with an entangled state of the remaining two qubits.

We give a few examples of the classification conditions. Consider a four-qubit state $\ket{\psi}$ and decompose it as in \eqref{eq:decomposition}. Let $\mathfrak{W}=\vspan{\ket{\phi_0},\ket{\phi_1}}$. Then:
\begin{itemize}
 \item If there is a spanning set for $\mathfrak{W}$ that contains two fully separable states, $\ket{\psi}$ is in the class $\mathfrak{W}_{000,000}$.
 \item If there is a spanning set containing one fully separable state and one bipartite separable state and there are no spanning sets containing two fully separable states, $\ket{\psi}$ is in the class $\mathfrak{W}_{000,0\Psi}$.
 \item If there is a spanning set containing one fully separable state and one GHZ state and there are no spanning sets containing two separable states of which at least one is fully separable, $\ket{\psi}$ is in the class $\mathfrak{W}_{000,GHZ}$.
 \item If there is a spanning set containing one fully separable state and one W state and there are no spanning sets containing a fully separable state together with a non-W type state, then $\ket{\psi}$ is in the class $\mathfrak{W}_{000,W}$.
 \item If there is a spanning set containing two bipartite separable states and there are no fully separable states in $\mathfrak{W}$, then $\ket{\psi}$ is in a $\mathfrak{W}_{0\Psi,0\Psi}$ class.
 \item And so on.
\end{itemize}
Always picking the spanning set with the `lowest' entanglement makes the classification unique: if e.g.\ some subspace $\mathfrak{W}$ has a spanning set containing two linearly-independent fully separable states, then there are certainly also spanning sets containing entangled states. Hence without the `lowest entanglement' criterion, many states would fall into multiple entanglement classes.

\section{Equations for the classification of three-qubit entanglement\label{s:equational_classification}}

For four or more qubits, the classification (or rather: the grouping of entanglement classes into superclasses) is not unique.
Different classification schemes can thus not easily be combined or interchanged.
That problem does not arise for three-qubit states: there is only a small finite number of entanglement classes -- fully separable states, three types of bipartite separable states, GHZ and W states -- whose definitions are generally accepted.
We are thus free to choose any method for identifying the entanglement class of a three-qubit state.
Hence, we use the following equational method derived by Li et al.\ \cite{li_simple_2006}, which we find more straightforward than the classification method in terms of coefficient matrix ranks from \cite{lamata_inductive_2006}.

This equational method works as follows.
Consider a three-qubit state expressed in the computational basis as:
\begin{multline}
 a_0\ket{000} + a_1\ket{001} + a_2\ket{010} + a_3\ket{011} + a_4\ket{100} \\ + a_5\ket{101} + a_6\ket{110} + a_7\ket{111},
\end{multline}
where $a_0,a_1,\ldots,a_7\in\CC$ are not all zero. 
This state is in the GHZ SLOCC class if and only if the following is non-zero:
\begin{equation}\label{eq:GHZ_criterion}
 (a_0a_7 - a_2a_5 + a_1a_6 - a_3a_4)^2 - 4(a_2a_4-a_0a_6)(a_3a_5-a_1a_7)
\end{equation}
The state is in the W SLOCC class if and only if \eqref{eq:GHZ_criterion} is zero and furthermore:
\begin{align}
 &(a_0a_3\neq a_1a_2 \vee a_5a_6\neq a_4a_7) \nonumber \\
 \wedge &(a_1a_4\neq a_0a_5 \vee a_3a_6\neq a_2a_7) \nonumber \\ \wedge &(a_3a_5\neq a_1a_7 \vee a_2a_4 \neq a_0a_6). \label{eq:W_criterion}
\end{align}
We sometimes refer to the formula \eqref{eq:W_criterion} as the `W conditions'.

In all other cases, the state is not genuinely three-partite entangled: all three clauses are false for a fully product state, for bipartite states, one of the clauses is satisfied and two are not \cite{li_simple_2006}.

This classification is basis-independent to some degree: as SLOCC transformations do not change the entanglement class of a state, the same equations hold when the state is expressed in any basis that arises from the computational basis via SLOCC, i.e.\ via applying an invertible 2 by 2 matrix to each qubit.

\section{The potentially empty four-qubit entanglement classes\label{s:results}}

In \cite{lamata_inductive_2007}, the inductive entanglement classification is applied to four-qubit states. A priori, and up to qubit permutations, there are $10$ potential entanglement classes, given that there are four entanglement classes of three-qubit states: $\mathfrak{W}_{000,000}$, $\mathfrak{W}_{000,0\Psi}$, $\mathfrak{W}_{000,GHZ}$, $\mathfrak{W}_{000,W}$, $\mathfrak{W}_{0\Psi,0\Psi}$, $\mathfrak{W}_{0\Psi,GHZ}$,  $\mathfrak{W}_{0\Psi,W}$, $\mathfrak{W}_{GHZ, GHZ}$, $\mathfrak{W}_{GHZ,W}$, and $\mathfrak{W}_{W,W}$.  

Lamata et al.\ consider these classes one-by-one and discard three of them as being empty: $\mathfrak{W}_{0\Psi,W}$, $\mathfrak{W}_{GHZ, GHZ}$ \footnote{Lamata et al.\ pick $\mathfrak{W}_{GHZ,W}$ as the default class instead of $\mathfrak{W}_{GHZ, GHZ}$ even though usually GHZ is ranked before W.}, and $\mathfrak{W}_{W,W}$. On the other hand, they split the class of states where the spanning set consists of two bipartite entangled states according to whether the two bipartitions are the same or different, resulting in $\mathfrak{W}_{0_k\Psi,0_k\Psi}$ and $\mathfrak{W}_{0_i\Psi,0_j\Psi}$. This gives a total of eight classes of genuine four-qubit entanglement. Additionally, there are also classes of four-qubit states that do not exhibit any four-partite entanglement; these are ignored here.

We re-analyse the three discarded classes using the equational classification criteria by Li et al.\ \cite{li_simple_2006} (cf.\ Section \ref{s:equational_classification}). With these methods, we show that, while $\mathfrak{W}_{GHZ, GHZ}$ is indeed empty, there are in fact states that belong into $\mathfrak{W}_{0\Psi,W}$ or $\mathfrak{W}_{W,W}$, respectively, according to the inductive classification.

\subsection{The entanglement class $\mathfrak{W}_{0_k\Psi,W}$}

The entanglement class labelled $\mathfrak{W}_{0_k\Psi,W}$ in the naming scheme of \cite{lamata_inductive_2007}, is the class containing states:
\begin{equation}
 \ket{\psi} = \ket{v_0}\ket{\phi_0}+\ket{v_1}\ket{\phi_1},
\end{equation}
where $\vspan{\ket{\phi_0},\ket{\phi_1})}$ contains no vectors of type $000$, exactly one vector of type $0\Psi$, and no vectors of GHZ type.

Up to permutations of the last three qubits, a generic representative of this class can be written as:
\begin{equation}
 \ket{\phi}\ket{\varphi_1}\ket{\Psi_1} + \ket{\bar{\phi}}\left(\ket{\varphi_2\psi_2\bar{\theta}_2}+\ket{\varphi_2\bar{\psi}_2\theta_2}+\ket{\bar{\varphi}_2\psi_2\theta_2}\right).
\end{equation}
Here, lowercase Greek letters label single-qubit states. Throughout, states do not need to be normalised (though they may not have norm 0). Overbars denote linear independence, i.e.\ $\ket{\phi}$ and $\ket{\bar{\phi}}$ are linearly independent, as are $\ket{\varphi_2}$ and $\ket{\bar{\varphi}_2}$, and so on. Different indices denote states that may or may not be linearly independent, e.g.\ $\ket{\varphi_1}$ may be linearly dependent on $\ket{\varphi_2}$ or $\ket{\bar{\varphi}_2}$, or neither. The state $\ket{\Psi_1}$ is an entangled two-qubit state.

To simplify the representative state, we can apply a SLOCC operation that maps $\left\{\ket{\phi},\ket{\bar{\phi}}\right\}$ to the computational basis, and similarly all pairs of linearly independent states with index 2. This yields:
\begin{multline}
 \ket{0}\ket{\varphi}\ket{\Psi} + \ket{1}\left(\ket{001}+\ket{010}+\ket{100}\right) \\
 = \ket{0}\ket{\varphi}\ket{\Psi} + \ket{1}\ket{W},
\end{multline}
where $\ket{\varphi}$ is the result of applying the SLOCC operation described above to $\ket{\varphi_1}$, and similarly for $\ket{\Psi}$.
Now $\ket{\varphi}$ and $\ket{\Psi}$ can be expressed in the computational basis as $\ket{\varphi}=\varphi_0\ket{0}+\varphi_1\ket{1}$ and:
\begin{equation}
 \ket{\Psi}=\Psi_{00}\ket{00}+\Psi_{01}\ket{01}+\Psi_{10}\ket{10}+\Psi_{11}\ket{11},
\end{equation}
where entanglement of $\ket{\Psi}$ implies that $\Psi_{00}\Psi_{11}-\Psi_{01}\Psi_{10}\neq 0$. Furthermore, $\varphi_0$ and $\varphi_1$ cannot both be zero.

Lamata et al.\ claim that $\vspan{\ket{\varphi\Psi}, \ket{W}}$ always contains a GHZ-type vector and that therefore the $\mathfrak{W}_{0_k\Psi,W}$ class is empty, as any state with a spanning set of type $0\Psi,W$ also has one of type ${0\Psi,GHZ}$ and thus falls into the $\mathfrak{W}_{0_k\Psi,GHZ}$ class by the ordering of the classes \cite{lamata_inductive_2007}. We show by analysis of the different combinations of parameter values that this is not correct and furthermore identify a canonical state for the $\mathfrak{W}_{0_k\Psi,W}$ class.

An arbitrary state in $\vspan{\ket{\varphi\Psi}, \ket{W}}$ can be written as $x\ket{\varphi\Psi} + y\ket{W}$ for some $x,y\in\CC$. Expanding this yields:
\begin{multline}
 x\varphi_0\Psi_{00}\ket{000} + (x\varphi_0\Psi_{01}+y)\ket{001} + (x\varphi_0\Psi_{10}+y)\ket{010} \\
 + x\varphi_0\Psi_{11}\ket{011} + (x\varphi_1\Psi_{00}+y)\ket{100} + x\varphi_1\Psi_{01}\ket{101} \\
 + x\varphi_1\Psi_{10}\ket{110} + x\varphi_1\Psi_{11}\ket{111}.
\end{multline}

Now, using \eqref{eq:GHZ_criterion}, we find that this state is in the GHZ SLOCC class if and only if:
\begin{widetext}
\begin{multline} 
 \Big(\left(\varphi_0^2\Psi_{11}^2+2\varphi_0\varphi_1\Psi_{01}\Psi_{11}+2\varphi_0\varphi_1\Psi_{10}\Psi_{11}+\varphi_1^2\Psi_{01}^2-2\varphi_1^2\Psi_{01}\Psi_{10}+\varphi_1^2\Psi_{10}^2+4\varphi_1^2\Psi_{00}\Psi_{11}\right)x + 4\varphi_1\Psi_{11} y\Big)xy^2 \neq 0.
\end{multline}
\end{widetext}
There are no GHZ type states in the subspace if and only if this polynomial is zero for all values of $x$ and $y$, i.e.\ if and only if:
\begin{multline}\label{eq:GHZ_polynomial}
 0 = \varphi_0^2\Psi_{11}^2+2\varphi_0\phi_1\Psi_{01}\Psi_{11}+2\varphi_0\varphi_1\Psi_{10}\Psi_{11}+\varphi_1^2\Psi_{01}^2 \\ -2\varphi_1^2\Psi_{01}\Psi_{10}+\varphi_1^2\Psi_{10}^2+4\varphi_1^2\Psi_{00}\Psi_{11}
\end{multline}
and also:
\begin{equation}\label{eq:GHZ_polynomial2}
 4\varphi_1\Psi_{11}=0.
\end{equation}

For each case in which the subspace contains no GHZ vectors, we furthermore need to check that it contains no separable vectors other than $\ket{\varphi\Psi}$. To exclude $\ket{\varphi\Psi}$, we assume that $y\neq 0$, and, for simplicity, rescale so that $y=1$.
We distinguish cases according to the solutions of \eqref{eq:GHZ_polynomial2}.

\subsubsection{Case $\varphi_1=0$\label{it:Psi_11=phi_1=0}}

If $\varphi_1=0$, then we must also have $\varphi_0\neq 0$. In this case, \eqref{eq:GHZ_polynomial} reduces to $\varphi_0^2\Psi_{11}^2=0$. As $\varphi_0$ cannot vanish, we must have $\Psi_{11}=0$ and hence $\Psi_{01}\Psi_{10}\neq 0$ by entanglement of $\ket{\Psi}$. The W conditions \eqref{eq:W_criterion} become:
\begin{multline}
 \left( 0 \neq (\varphi_0\Psi_{01}x+1)(\varphi_0\Psi_{10}x+1) \right) \\ \wedge \left(\varphi_0\Psi_{01}x+1 \neq 0\right)  \wedge \left(\varphi_0\Psi_{10}x+1 \neq 0\right).
\end{multline}
All parameters appearing in those inequalities are non-zero, thus there are always separable states in the subspace: set $x = -1/(\varphi_0\Psi_{01})$ or $x = -1/(\varphi_0\Psi_{10})$.

\subsubsection{Case $\Psi_{11}=0$}
 
If $\Psi_{11}=0$, then we must also have $\Psi_{01}\Psi_{10}\neq 0$. In this case, \eqref{eq:GHZ_polynomial} reduces to $\varphi_1^2(\Psi_{01}-\Psi_{10})^2 =0$. Hence there are two subcases.
  \begin{enumerate}
   \item $\varphi_1=0$, which implies $\varphi_0\neq 0$.  This brings us back to the case considered in Section \ref{it:Psi_11=phi_1=0} above.
   \item $\Psi_{01}=\Psi_{10}$: under this assumption, the state is in the W SLOCC class if:
    \begin{multline}\label{eq:0psi_w}
     \left( \left((0\neq (\varphi_0\Psi_{01}x+1)^2\right) \vee \left(\varphi_1^2\Psi_{01}^2x^2\neq 0\right)\right) \\ \wedge \left((\varphi_0\Psi_{01}+\varphi_1\Psi_{00})x+1\neq 0\right).
    \end{multline}
    The last inequality represents two clauses in the original set of three, which have become identical under the current choice of parameter values.
    Now, the following cases occur:
    \begin{itemize}
     \item\label{it:phi_1=0} If $\varphi_1=0\neq\varphi_0$, there exists a $000$ type state in the subspace: setting $x=-1/(\varphi_0\Psi_{01})$ makes all the inequalities false.
     \item If $\varphi_1\neq 0$ and $\varphi_0\Psi_{01}+\varphi_1\Psi_{00}\neq 0$, there is a separable state in the space, which can be constructed by setting $x = -1/(\varphi_0\Psi_{01}+\varphi_1\Psi_{00})$. (Recall that the last inequality represents two clauses of the original set.)
     \item If $\varphi_1\neq 0$ and $\varphi_0\Psi_{01}+\varphi_1\Psi_{00}=0$, any state in the subspace (other than $\ket{\varphi\Psi}$) is in the W class. To see this, note that $\varphi_0\Psi_{01}+\varphi_1\Psi_{00}=0$ implies that the last inequality in \eqref{eq:0psi_w} is always satisfied. The first two inequalities would both need to be false simultaneously for the state not to be in the W class. But the second inequality is false only for $x=0$, for which the first inequality is satisfied. Hence states of this form are classified into $\mathfrak{W}_{0_k\Psi,W}$ by the inductive scheme.
    \end{itemize}
  \end{enumerate}
This concludes the analysis of all cases in which $\vspan{\ket{\varphi\Psi}, \ket{W}}$ contains no GHZ-type states.

\subsubsection{The canonical state for $\mathfrak{W}_{0_k\Psi,W}$}

From the above, the canonical state for $\mathfrak{W}_{0_k\Psi,W}$ satisfies $\Psi_{11}=0$, $\Psi_{01}=\Psi_{10}$, $\varphi_1\neq 0$ and $\varphi_0\Psi_{01}+\varphi_1\Psi_{00}=0$. Since $\varphi_1\neq 0$, we can write $\Psi_{00}=-\varphi_0\Psi_{01}/\varphi_1$. Then the canonical generator is:
\begin{equation}
 (\varphi_0\ket{0}+\varphi_1\ket{1})\left(-\frac{\varphi_0\Psi_{01}}{\varphi_1}\ket{00}+\Psi_{01}\ket{\Psi^+}\right),
\end{equation}
where $\ket{\Psi^{+}}=\ket{01}+\ket{10}$. Let $\lambda=\varphi_0/\varphi_1$, and the generator becomes:
\begin{equation}
 \varphi_1\Psi_{01} \left(\lambda\ket{0}+\ket{1}\right) \left(-\lambda\ket{00} + \ket{\Psi^+}\right).
\end{equation}
From the previous conditions, we must have $\lambda,\varphi_1,\Psi_{01}\in\CC$ and $\varphi_1\Psi_{01}\neq 0$. The canonical state is:
\begin{equation}
 \varphi_1\Psi_{01}\ket{0}\left(\lambda\ket{0}+\ket{1}\right) \left(-\lambda\ket{00} + \ket{\Psi^+}\right) + \ket{1}\ket{W},
\end{equation}
Note that the non-zero factor $\varphi_1\Psi_{01}$ can be removed by a SLOCC operation on the first qubit. Then the canonical state becomes:
\begin{equation}
 \ket{0}\left(\lambda\ket{0}+\ket{1}\right) \left(-\lambda\ket{00} + \ket{\Psi^+}\right) + \ket{1}\ket{W},
\end{equation}
where $\lambda$ is arbitrary.

\subsection{The entanglement class $\mathfrak{W}_{GHZ,GHZ}$}

Lamata et al.\ consider the class $\mathfrak{W}_{GHZ,W}$ before $\mathfrak{W}_{GHZ,GHZ}$, i.e.\ if a state qualifies for both, it is considered to be in $\mathfrak{W}_{GHZ,W}$. This means that a state is in the $\mathfrak{W}_{GHZ,GHZ}$ if and only if $\vspan{\ket{\phi_0},\ket{\phi_1}}$ contains only GHZ type vectors, where the state has been decomposed as in \eqref{eq:decomposition}.
Now, Lamata et al.\ argue that the span of two GHZ type vectors always contains a W state or a separable state and that therefore $\mathfrak{W}_{GHZ,GHZ}$ is empty \cite{lamata_inductive_2007}. To ensure that our revisions of the classification are complete, we confirm this result using the methods from \cite{li_simple_2006}.

A generic representative of $\mathfrak{W}_{GHZ,GHZ}$ can be written as:
\begin{equation}
 \ket{\phi}\left(\ket{\varphi_1\psi_1\theta_1}+\ket{\bar{\varphi}_1\bar{\psi}_1\bar{\theta}_1}\right) + \ket{\bar{\phi}}\left(\ket{\varphi_2\psi_2\theta_2}+\ket{\bar{\varphi}_2\bar{\psi}_2\bar{\theta}_2}\right)
\end{equation}
with the same conventions as before. Via a SLOCC operation, this state can be transformed to:
\begin{multline}
 \ket{0}\left(\ket{\varphi\psi\theta}+\ket{\bar{\varphi}\bar{\psi}\bar{\theta}}\right) + \ket{1}\left(\ket{000}+\ket{111}\right) \\ 
 = \ket{0}\left(\ket{\varphi\psi\theta}+\ket{\bar{\varphi}\bar{\psi}\bar{\theta}}\right) + \ket{1}\ket{GHZ},
\end{multline}
where $\ket{\varphi}$ is the result of applying the SLOCC operation to $\ket{\varphi_1}$, and similarly for the other variable states.

Rather than expanding each single-qubit state in the computational basis (leading to 12 parameters and three linear-independence conditions), it will be easier to expand the full state $\ket{\varphi\psi\theta}+\ket{\bar{\varphi}\bar{\psi}\bar{\theta}}$ instead, which only requires eight parameters and the GHZ condition. Thus let:
\begin{multline}
 \ket{\varphi\psi\theta}+\ket{\bar{\varphi}\bar{\psi}\bar{\theta}} \\
 = a_0\ket{000} + a_1\ket{001} + a_2\ket{010} + a_3\ket{011} \qquad \\ + a_4\ket{100} + a_5\ket{101} + a_6\ket{110} + a_7\ket{111},
\end{multline}
where $a_0,\ldots,a_7\in\CC$ satisfy:
\begin{multline}
 0\neq (a_0a_7 - a_2a_5 + a_1a_6 - a_3a_4)^2 \\ - 4(a_2a_4-a_0a_6)(a_3a_5-a_1a_7).
\end{multline}
A general element of $\vspan{\ket{\varphi\psi\theta}+\ket{\bar{\varphi}\bar{\psi}\bar{\theta}},\ket{GHZ}}$ then has the form:
\begin{equation}
 x\left(\ket{\varphi\psi\theta}+\ket{\bar{\varphi}\bar{\psi}\bar{\theta}}\right) + y \ket{GHZ},
\end{equation}
or, equivalently:
\begin{multline}
 (a_0x+y)\ket{000} + a_1x\ket{001} + a_2x\ket{010} + a_3x\ket{011} \\
 + a_4x\ket{100} + a_5x\ket{101} + a_6x\ket{110} + (a_7x+y)\ket{111},
\end{multline}
where $x,y\in\CC$.
For a state to be in the class $\mathfrak{W}_{GHZ,GHZ}$, all non-vanishing elements of the subspace must be of GHZ type, i.e. we need the polynomial:
\begin{widetext}
\begin{multline}
 \left(a_{3}^{2} a_{4}^{2} - 2 a_{2} a_{3} a_{4} a_{5} + a_{2}^{2} a_{5}^{2} - 2 a_{1} a_{3} a_{4} a_{6} - 2 a_{1} a_{2} a_{5} a_{6} + 4 a_{0} a_{3} a_{5} a_{6} + a_{1}^{2} a_{6}^{2} + 4 a_{1} a_{2} a_{4} a_{7} - 2 a_{0} a_{3} a_{4} a_{7} - 2 a_{0} a_{2} a_{5} a_{7} \right. \\
 \left. - 2 a_{0} a_{1} a_{6} a_{7} + a_{0}^{2} a_{7}^{2} \right) x^{4} + \left( 4 a_{1} a_{2} a_{4} - 2 a_{0} a_{3} a_{4} - 2 a_{0} a_{2} a_{5} - 2 a_{0} a_{1} a_{6} + 4 a_{3} a_{5} a_{6} + 2 a_{0}^{2} a_{7} - 2 a_{3} a_{4} a_{7} - 2 a_{2} a_{5} a_{7} \right. \\ 
 \left. - 2 a_{1} a_{6} a_{7} + 2 a_{0} a_{7}^{2} \right) x^{3} y + \left( a_{0}^{2} - 2 a_{3} a_{4} - 2 a_{2} a_{5} - 2 a_{1} a_{6} + 4 a_{0} a_{7} + a_{7}^{2} \right) x^{2} y^{2} + 2 \left( a_{0} + a_{7} \right) x y^{3} + y^{4}
\end{multline}
\end{widetext}
to be non-zero whenever at least one of $x,y$ is non-zero.
This means that, given any fixed non-zero value for $x$ (or $y$), the remaining polynomial in the other variable must have no roots.
A polynomial has no roots over $\CC$ only if it is equal to some non-zero constant.
Hence for any non-zero $x$, the polynomial must be independent of $y$.
That cannot be achieved here for any choice of $a_0,\ldots,a_7$ as the coefficient of $y^4$ is independent of the parameters and of $x$.
Hence the subspace spanned by two GHZ states always contains non-GHZ states and the class $\mathfrak{W}_{GHZ,GHZ}$ is indeed empty.

\subsection{The entanglement class $\mathfrak{W}_{W,W}$}

Finally, consider the class $\mathfrak{W}_{W,W}$, i.e.\ the class containing states:
\begin{equation}
 \ket{\psi} = \ket{v_0}\ket{\phi_0}+\ket{v_1}\ket{\phi_1},
\end{equation}
where $\vspan{\ket{\phi_0},\ket{\phi_1}}$ contains only W type vectors.

A generic representative of this class can be written as:
\begin{multline}
 \ket{\phi}\left(\ket{\varphi_1\psi_1\bar{\theta}_1}+\ket{\varphi_1\bar{\psi}_1\theta_1}+\ket{\bar{\varphi}_1\psi_1\theta_1}\right) \\
 + \ket{\bar{\phi}}\left(\ket{\varphi_2\psi_2\bar{\theta}_2}+\ket{\varphi_2\bar{\psi}_2\theta_2}+\ket{\bar{\varphi}_2\psi_2\theta_2}\right).
\end{multline}
As before, overbars denote linear independence. States with different indices may or may not be linearly dependent, as long as the two W type states are linearly independent from each other.

To simplify this representative state, apply a SLOCC operation that maps $\left\{\ket{\phi},\ket{\bar{\phi}}\right\}$ to the computational basis, and similarly all pairs of linearly independent states with index 2. This yields:
\begin{multline}
 \ket{0}\left(\ket{\varphi\psi\bar{\theta}}+\ket{\varphi\bar{\psi}\theta}+\ket{\bar{\varphi}\psi\theta}\right) \\
 + \ket{1}\left(\ket{001}+\ket{010}+\ket{100}\right),
\end{multline}
where $\ket{\varphi}$ is the result of applying the SLOCC operation described above to $\ket{\varphi_1}$, and similarly for the other variable states.
Rather than expanding each single-qubit state in the computational basis, it is easier to expand the full state and use the W conditions from Section \ref{s:equational_classification}. Thus let:
\begin{multline}
 \ket{\varphi\psi\bar{\theta}}+\ket{\varphi\bar{\psi}\theta}+\ket{\bar{\varphi}\psi\theta} \\
 = a_0\ket{000} + a_1\ket{001} + a_2\ket{010} + a_3\ket{011} \qquad \\
 + a_4\ket{100} + a_5\ket{101} + a_6\ket{110} + a_7\ket{111},
\end{multline}
where $a_0,\ldots,a_7\in\CC$ satisfy:
\begin{multline}
 0 = (a_0a_7 - a_2a_5 + a_1a_6 - a_3a_4)^2 \\ - 4(a_2a_4-a_0a_6)(a_3a_5-a_1a_7)
\end{multline}
and:
\begin{align}
 &(a_0a_3\neq a_1a_2 \vee a_5a_6\neq a_4a_7) \nonumber \\
 \wedge &(a_1a_4\neq a_0a_5 \vee a_3a_6\neq a_2a_7) \nonumber \\
 \wedge &(a_3a_5\neq a_1a_7 \vee a_2a_4 \neq a_0a_6).
\end{align}

A general element of:
\begin{equation}
 \vspan{\ket{\varphi\psi\bar{\theta}}+\ket{\varphi\bar{\psi}\theta}+\ket{\bar{\varphi}\psi\theta},\ket{W}}
\end{equation}
has the form:
\begin{equation}
 x \left( \ket{\varphi\psi\bar{\theta}} + \ket{\varphi\bar{\psi}\theta} + \ket{\bar{\varphi}\psi\theta} \right) + y \ket{W},
\end{equation}
or:
\begin{multline}
  a_0x\ket{000} + (a_1x+y)\ket{001} + (a_2x+y)\ket{010} + a_3x\ket{011} \\
 + (a_4x+y)\ket{100} + a_5x\ket{101} + a_6x\ket{110} + a_7x\ket{111}
\end{multline}
where $x,y\in\CC$.
We need all non-vanishing elements of this subspace to be of W type (the case $y=0$ automatically includes $\ket{\varphi\psi\bar{\theta}}+\ket{\varphi\bar{\psi}\theta}+\ket{\bar{\varphi}\psi\theta}$ itself). By \eqref{eq:GHZ_criterion}, this requires the following polynomial in $x$ and $y$ to be identically zero:
\begin{widetext}
\begin{multline}\label{eq:GHZ_poly_ww}
 \left( a_{3}^{2} a_{4}^{2} - 2 a_{2} a_{3} a_{4} a_{5} + a_{2}^{2} a_{5}^{2} - 2 a_{1} a_{3} a_{4} a_{6} - 2 a_{1} a_{2} a_{5} a_{6} + 4 a_{0} a_{3} a_{5} a_{6} + a_{1}^{2} a_{6}^{2} + 4 a_{1} a_{2} a_{4} a_{7} - 2 a_{0} a_{3} a_{4} a_{7} \right. \\
 \left. - 2 a_{0} a_{2} a_{5} a_{7} - 2 a_{0} a_{1} a_{6} a_{7} + a_{0}^{2} a_{7}^{2} \right) x^4 + \left( 2 a_{3}^{2} a_{4} - 2 a_{2} a_{3} a_{5} - 2 a_{3} a_{4} a_{5} + 2 a_{2} a_{5}^{2} - 2 a_{1} a_{3} a_{6} - 2 a_{3} a_{4} a_{6} \right. \\
 \left. - 2 a_{1} a_{5} a_{6} - 2 a_{2} a_{5} a_{6} + 2 a_{1} a_{6}^{2} + 4 a_{1} a_{2} a_{7} - 2 a_{0} a_{3} a_{7} + 4 a_{1} a_{4} a_{7} + 4 a_{2} a_{4} a_{7} - 2 a_{0} a_{5} a_{7}  - 2 a_{0} a_{6} a_{7} \right) x^3 y \\
 + \left( a_{3}^{2} - 2 a_{3} a_{5} + a_{5}^{2} - 2 a_{3} a_{6} - 2 a_{5} a_{6} + a_{6}^{2} + 4 a_{1} a_{7} + 4 a_{2} a_{7} + 4 a_{4} a_{7} \right) x^2 y^2 + 4 a_{7} x y^3.
\end{multline}
\end{widetext}
Additionally, by \eqref{eq:W_criterion}, it requires the following three statements to be true whenever $x,y$ are not both zero:
\begin{multline}\label{eq:W_ww1}
 \left( y^2 + (a_1+a_2)xy + (a_{1} a_{2} - a_{0} a_{3})x^2 \neq 0 \right) \\
 \vee \left( a_7 xy + (a_4a_7 - a_5a_6)x^2 \neq 0 \right)
\end{multline}
\begin{multline}\label{eq:W_ww2}
 \left( y^2 + (a_1+a_4)xy + (a_1a_4-a_0a_5)x^2 \neq 0 \right) \\
 \vee \left( a_7 xy + (a_2a_7 - a_3a_6)x^2 \neq 0 \right)
\end{multline}
\begin{multline}\label{eq:W_ww3}
 \left( a_7xy + (a_1a_7 - a_3a_5)x^2 \neq 0 \right) \\
 \vee \left( y^2 + (a_2+a_4)xy + (a_2a_4 - a_0a_6)x^2 \neq 0 \right).
\end{multline}

The GHZ polynomial \eqref{eq:GHZ_poly_ww} is identically zero if and only if each coefficient is zero. For the coefficient of $xy^3$ this implies $a_7=0$. Given that assumption, the coefficient of $x^2y^2$ vanishes if:
\begin{equation}
 a_3^3-2a_3a_5-2a_3a_6+a_5^2-2a_5a_6+a_6^2=0.
\end{equation}
There are several cases.

\subsubsection{Case $a_6=0$ and $a_5=a_3$}

With $a_6=0$ and $a_5=a_3$, the GHZ polynomial reduces to $(a_2-a_4)^2a_3^2 x^4$. Hence there are two subcases:
\begin{enumerate}
   \item $a_3=0$, in which case the W conditions \eqref{eq:W_ww1}--\eqref{eq:W_ww3} become:
   \begin{align}
    &\left( 0\neq (a_1x+y)(a_2x+y) \right) \nonumber \\
    \wedge &\left( (a_1x+y)(a_4x+y) \neq 0 \right) \nonumber \\
    \wedge &\left( (a_2x+y)(a_4x+y) \neq 0 \right)
   \end{align}
   As the $y=0$ case must be a W type state, this implies $a_1,a_2,a_4\neq 0$. But then the subspace always contains separable states by setting $y=-a_1x$ or $y=-a_2x$ or $y=-a_4x$ for any non-zero $x$.
   \item $a_3\neq 0$ and $a_4=a_2$, in which case the W conditions become:
   \begin{multline}
    \left( (a_1a_2-a_0a_3)x^2 + (a_1+a_2)xy + y^2 \neq 0 \right) \\
    \wedge \left( \left( a_3^2 x^2\neq 0 \right)
    \vee \left( (a_2x+y)^2\neq 0 \right) \right),
   \end{multline}
   where the first inequality represents both \eqref{eq:W_ww1} and \eqref{eq:W_ww2}. Hence for the $y=0$ case to be a W type state, we need $a_0a_3\neq a_1a_2$. But then the subspace always contains separable states by setting $y$ to a root of $(a_{1} a_{2} - a_{0} a_{3})x^2 + (a_1+a_2)xy + y^2$ (where $x$ is considered as another parameter and we are free to assume it is non-zero).
\end{enumerate}
The cases $((a_5=0)\wedge (a_6=a_3))$ and $((a_3=0)\wedge (a_6=a_5))$ are analogous.

\subsubsection{Case $a_3,a_5,a_6\neq 0$ and $a_6=a_3+a_5\pm 2\sqrt{a_3a_5}$}

First, note that $a_3a_5\pm a_3\sqrt{a_3a_5}=0$ would imply that at least one of $a_3$, $a_5$ and $a_6$ is zero. Hence, in the current case we must have $a_3a_5\pm a_3\sqrt{a_3a_5}\neq 0$. Thus, for the coefficient of $x^3y$ to vanish, we require:
\begin{equation}
   a_4 = \frac{2a_1a_3a_5-a_2a_3a_5\pm\sqrt{a_3a_5}(a_1a_3+a_1a_5-a_2a_5)}{a_3a_5\pm a_3\sqrt{a_3a_5}},
\end{equation}
and this expression is well-defined. Then the GHZ polynomial becomes:
\begin{widetext}
  \begin{equation}
   \frac{{\left(a_{1}^{2} - 2 a_{1} a_{2} + a_{2}^{2} + 4 a_{0} a_{3}\right)} {\left(a_{3}^{2} + 6 a_{3} a_{5} + a_{5}^{2} \pm 4 \sqrt{a_{3} a_{5}} a_{3} \pm 4 \sqrt{a_{3} a_{5}} a_{5}\right)} a_{5}^{2} x^4}{{\left(a_{5} \pm \sqrt{a_{3} a_{5}}\right)}^{2}}.
  \end{equation}
\end{widetext}
Again, this is well-defined as $a_5\pm\sqrt{a_3a_5}=0$ implies either $a_5=0$, or $a_3=a_5$ and $a_6=0$; so under the assumptions of the current case, $a_5\pm\sqrt{a_3a_5}$ must always be non-zero.

There are two subcases.
\begin{enumerate}
      \item $a_{3}^{2} + 6 a_{3} a_{5} + a_{5}^{2} \pm 4 \sqrt{a_{3} a_{5}} a_{3} \pm 4 \sqrt{a_{3} a_{5}} a_{5}=0$. This implies:
       \begin{equation}
        \left(a_{3}^{2} + 6 a_{3} a_{5} + a_{5}^{2}\right)^2 = 16a_{3} a_{5}\left( a_{3}+a_{5}\right)^2,
       \end{equation}
       which is equivalent to $(a_3-a_5)^4=0$, so it implies $a_3=a_5$. Then, the equation $a_{3}^{2} + 6 a_{3} a_{5} + a_{5}^{2} \pm 4 \sqrt{a_{3} a_{5}} a_{3} \pm 4 \sqrt{a_{3} a_{5}} a_{5}=0$ becomes:
       \begin{equation}
        8a_3\left(a_3\pm\sqrt{a_3^2}\right)=0.
       \end{equation}
       We assumed $a_3\neq 0$, so this can only be satisfied if the square root function and sign are such that $a_3\pm\sqrt{a_3^2}=0$. But $a_3=a_5$ and $a_3\pm\sqrt{a_3^2}=0$ together imply that $a_3+a_5\pm 2\sqrt{a_3a_5}=0$, which contradicts the assumption that $a_6\neq 0$. Hence this case cannot happen.
       
      \item $a_{1}^{2} - 2 a_{1} a_{2} + a_{2}^{2} + 4 a_{0} a_{3}=0$, i.e. $a_0 = -(a_1-a_2)^2/(4a_3)$.
       In this case, the subspace contains no GHZ states (by construction). The W conditions \eqref{eq:W_ww1}--\eqref{eq:W_ww3} become:
       \begin{multline}
        \left(-\frac{1}{4}\left((a_1+a_2)x + 2y\right)^2 \neq 0 \right) \\
        \vee \left( \left(a_3+a_5\pm 2\sqrt{a_3a_5} \right) a_5 x^2 \neq 0 \right),
       \end{multline}
       \begin{equation}
        \left( P(x,y) \neq 0 \right) \vee \left( \left(a_3+a_5\pm 2\sqrt{a_3a_5} \right) a_3 x^2 \neq 0 \right),
       \end{equation}
       and
       \begin{equation}
        \left( a_3 a_5 x^2 \neq 0 \right) \vee \left( Q(x,y) \neq 0 \right),
       \end{equation}
       where:
       \begin{widetext}
       \begin{multline}
        P(x,y) = \left( 4 a_{3} {\left(a_{5} \pm \sqrt{a_{3} a_{5}}\right)} \right)^{-1} 
        \Big( 4 \left( 3 a_{1} a_{3} a_{5} - a_{2} a_{3} a_{5} \pm 2 \sqrt{a_{3} a_{5}} a_{1} a_{3} \pm \sqrt{a_{3} a_{5}} a_{1} a_{5} \mp \sqrt{a_{3} a_{5}} a_{2} a_{5} \right) x y \\
        + 4 a_3 \left( a_{5} \pm \sqrt{a_{3} a_{5}} \right) y^{2} + \left( 8 a_{1}^{2} a_{3} a_{5} - 4 a_{1} a_{2} a_{3} a_{5} + a_{1}^{2} a_{5}^{2} - 2 a_{1} a_{2} a_{5}^{2} + a_{2}^{2} a_{5}^{2} \pm 4 \sqrt{a_{3} a_{5}} a_{1}^{2} a_{3} \right. \\
        \left. \pm 5 \sqrt{a_{3} a_{5}} a_{1}^{2} a_{5} \mp 6 \sqrt{a_{3} a_{5}} a_{1} a_{2} a_{5} \pm \sqrt{a_{3} a_{5}} a_{2}^{2} a_{5} \right) x^{2} \Big)
       \end{multline}
       and:
       \begin{multline}
        Q(x,y) = \left( 4 a_{3} {\left(a_{5} \pm \sqrt{a_{3} a_{5}}\right)} \right)^{-1} \Big( 4 \left( 2 a_{1} a_{3} a_{5} \pm \sqrt{a_{3} a_{5}} a_{1} a_{3} \pm \sqrt{a_{3} a_{5}} a_{2} a_{3} \pm \sqrt{a_{3} a_{5}} a_{1} a_{5} \mp \sqrt{a_{3} a_{5}} a_{2} a_{5} \right) x y \\
        + 4 a_{3} \left( a_{5} \pm \sqrt{a_{3} a_{5}} \right) y^{2} + \left( 3 a_{1}^{2} a_{3} a_{5} + 2 a_{1} a_{2} a_{3} a_{5} - a_{2}^{2} a_{3} a_{5} + a_{1}^{2} a_{5}^{2} - 2 a_{1} a_{2} a_{5}^{2} + a_{2}^{2} a_{5}^{2} \pm \sqrt{a_{3} a_{5}} a_{1}^{2} a_{3} \right. \\
        \left. \pm 2 \sqrt{a_{3} a_{5}} a_{1} a_{2} a_{3} \pm \sqrt{a_{3} a_{5}} a_{2}^{2} a_{3} \pm 3 \sqrt{a_{3} a_{5}} a_{1}^{2} a_{5} \mp 2 \sqrt{a_{3} a_{5}} a_{1} a_{2} a_{5} \mp \sqrt{a_{3} a_{5}} a_{2}^{2} a_{5} \right) x^2 \Big).
       \end{multline}
       \end{widetext}
       By assumption, $a_3,a_5$ and $a_3+a_5\pm 2\sqrt{a_3a_5}$ are non-zero, so for $x\neq 0$ the W conditions are satisfied. With $x=0\neq y$, the W conditions can be seen to reduce to just one inequality, $y^2\neq 0$, which is clearly satisfied (as it should be, since the $x=0$ state is $y\ket{W}$ and thus a W state by construction). This means that the subspace contains no separable states and states of this type belong in $\mathfrak{W}_{W,W}$ according to the classification.
       
       The canonical generator has the form:
       \begin{multline}
        -\frac{(a_1-a_2)^2}{4a_3} \ket{000} + a_{1} \ket{001} + a_{2} \ket{010} + a_{3} \ket{011} \\ + \frac{ (2 a_{1} - a_{2}) a_{3} a_{5} \pm {\left(a_{1} a_{3} + a_{1} a_{5} - a_{2} a_{5}\right)} \sqrt{a_{3} a_{5}}}{a_{3} a_{5} \pm \sqrt{a_{3} a_{5}} a_{3}} \ket{100} \\ + a_{5} \ket{101}
        + {\left(a_{3} + a_{5} \pm 2 \sqrt{a_{3} a_{5}}\right)} \ket{110},
       \end{multline}
       where $a_3,a_5$ are non-zero with $a_3+a_5\pm 2\sqrt{a_3a_5}$ also non-zero for the given choice of square root function and sign, and $a_1,a_2$ are arbitrary.
\end{enumerate}
This concludes the investigation of all potential members of $\mathfrak{W}_{W,W}$.

\subsubsection{The canonical state for $\mathfrak{W}_{W,W}$}

Given the canonical generator above, the canonical state for $\mathfrak{W}_{W,W}$ can be written (up to SLOCC) as:
\begin{widetext}
\begin{multline}
 \ket{0}\left(-\frac{(a_1-a_2)^2}{4a_3} \ket{000} + a_{1} \ket{001} + a_{2} \ket{010} + a_{3} \ket{011} + \frac{ (2 a_{1} - a_{2}) a_{3} a_{5} \pm {\left(a_{1} a_{3} + a_{1} a_{5} - a_{2} a_{5}\right)} \sqrt{a_{3} a_{5}}}{a_{3} a_{5} \pm \sqrt{a_{3} a_{5}} a_{3}} \ket{100} \right. \\
 \left. \vphantom{\frac{ (2 a_{1} - a_{2}) a_{3} a_{5} \pm {\left(a_{1} a_{3} + a_{1} a_{5} - a_{2} a_{5}\right)} \sqrt{a_{3} a_{5}}}{a_{3} a_{5} \pm \sqrt{a_{3} a_{5}} a_{3}}}+ a_{5} \ket{101} + {\left(a_{3} + a_{5} \pm 2 \sqrt{a_{3} a_{5}}\right)} \ket{110} \right) + \ket{1}\ket{W},
\end{multline}
where $a_3,a_5$ are non-zero with $a_3+a_5\pm 2\sqrt{a_3a_5}$ also non-zero for the given choice of square root function and sign, and $a_1,a_2$ are arbitrary.

In fact, we can remove one parameter via SLOCC. Note that:
\begin{multline}
 -\frac{(a_1-a_2)^2}{4a_3} \ket{000} + a_{1} \ket{001} + a_{2} \ket{010} + a_{3} \ket{011} + \frac{ (2 a_{1} - a_{2}) a_{3} a_{5} \pm {\left(a_{1} a_{3} + a_{1} a_{5} - a_{2} a_{5}\right)} \sqrt{a_{3} a_{5}}}{a_{3} a_{5} \pm \sqrt{a_{3} a_{5}} a_{3}} \ket{100} \\
 + a_{5} \ket{101} + {\left(a_{3} + a_{5} \pm 2 \sqrt{a_{3} a_{5}}\right)} \ket{110}
\end{multline}
is equal to:
\begin{multline}
 -\frac{(a_1-a_2)^2}{4a_3} \ket{000} + (a_2-a_1) \ket{010} + a_{3} \ket{011} + \frac{ (a_{1} - a_{2})\left( a_{3} a_{5} \pm a_{5}\sqrt{a_{3} a_{5}}\right)}{a_{3} a_{5} \pm a_{3} \sqrt{a_{3} a_{5}}} \ket{100} \\
 + a_{5} \ket{101}
 + {\left(a_{3} + a_{5} \pm 2 \sqrt{a_{3} a_{5}}\right)} \ket{110} + a_1\ket{W},
\end{multline}
where we have separated out a copy of $a_1\ket{W}$.
Now let $\mu=a_1-a_2$, then the canonical state can be transformed to:
\begin{multline}
 \ket{0}\left(-\frac{\mu^2}{4a_3} \ket{000} - \mu \ket{010} + a_{3} \ket{011} + \mu \frac{ a_5 \left(a_{3} \pm \sqrt{a_{3} a_{5}} \right)}{a_{3} \left( a_{5} \pm \sqrt{a_{3} a_{5}}\right)} \ket{100} + a_{5} \ket{101} + {\left(a_{3} + a_{5} \pm 2 \sqrt{a_{3} a_{5}}\right)} \ket{110} \right) + \ket{1}\ket{W}
\end{multline}
by a SLOCC operation on the first qubit.
\end{widetext}

\section{Conclusions}

In their inductive classification of four-qubit states, Lamata et al.\ discard three potential superclasses as not having any members, namely the ones labelled $\mathfrak{W}_{0_k\Psi,W}$, $\mathfrak{W}_{GHZ,GHZ}$ and $\mathfrak{W}_{W,W}$.
We show that, while they are correct in stating that $\mathfrak{W}_{GHZ,GHZ}$ is empty, the other two classes are in fact non-empty.

In particular, we find that the $\mathfrak{W}_{0_k\Psi,W}$ class consists of the following one-parameter family of states, up to SLOCC and permutations of the last three qubits:
\begin{equation}
 \ket{0} \left(\lambda\ket{0}+\ket{1}\right) \left(-\lambda\ket{00} + \ket{\Psi^+}\right)+\ket{1}\ket{W},
\end{equation}
where $\lambda\in\CC$.

Furthermore, we show that $\mathfrak{W}_{W,W}$ consists of the following three-parameter family of states, again up to SLOCC:
\begin{multline}
 \ket{0}\left(-\frac{\mu^2}{4a_3} \ket{000} - \mu \ket{010} + a_{3} \ket{011} \right. \\
 + \mu \frac{ a_5 \left(a_{3} \pm \sqrt{a_{3} a_{5}} \right)}{a_{3} \left( a_{5} \pm \sqrt{a_{3} a_{5}}\right)} \ket{100} + a_{5} \ket{101} \\
 \left.\vphantom{\frac{ a_5 \left(a_{3} \pm \sqrt{a_{3} a_{5}} \right)}{a_{3} \left( a_{5} \pm \sqrt{a_{3} a_{5}}\right)}} + {\left(a_{3} + a_{5} \pm 2 \sqrt{a_{3} a_{5}}\right)} \ket{110} \right) + \ket{1}\ket{W},
\end{multline}
where $a_3$ and $a_5$ are non-zero complex numbers satisfying $a_3+a_5\pm 2\sqrt{a_3a_5}\neq 0$ for the given choice of square root function and sign, and $\mu$ is an arbitrary complex number.

In the inductive classification scheme, there are hence ten (rather than eight) entanglement superclasses of four-qubit genuinely entangled states, with corresponding effects on the expected number of entanglement superclasses of five or more qubits.


\begin{acknowledgments}
 I would like to thank Ashley Montanaro and Lucas Lamata for helpful comments on earlier versions of this paper. The SageMath software system \cite{stein_sage_2016} was very useful for handling the symbolic algebra. I acknowledge funding from EPSRC.
\end{acknowledgments}

\bibliography{refs}

\begin{thebibliography}{8}%
\makeatletter
\providecommand \@ifxundefined [1]{%
 \@ifx{#1\undefined}
}%
\providecommand \@ifnum [1]{%
 \ifnum #1\expandafter \@firstoftwo
 \else \expandafter \@secondoftwo
 \fi
}%
\providecommand \@ifx [1]{%
 \ifx #1\expandafter \@firstoftwo
 \else \expandafter \@secondoftwo
 \fi
}%
\providecommand \natexlab [1]{#1}%
\providecommand \enquote  [1]{``#1''}%
\providecommand \bibnamefont  [1]{#1}%
\providecommand \bibfnamefont [1]{#1}%
\providecommand \citenamefont [1]{#1}%
\providecommand \href@noop [0]{\@secondoftwo}%
\providecommand \href [0]{\begingroup \@sanitize@url \@href}%
\providecommand \@href[1]{\@@startlink{#1}\@@href}%
\providecommand \@@href[1]{\endgroup#1\@@endlink}%
\providecommand \@sanitize@url [0]{\catcode `\\12\catcode `\$12\catcode
  `\&12\catcode `\#12\catcode `\^12\catcode `\_12\catcode `\%12\relax}%
\providecommand \@@startlink[1]{}%
\providecommand \@@endlink[0]{}%
\providecommand \url  [0]{\begingroup\@sanitize@url \@url }%
\providecommand \@url [1]{\endgroup\@href {#1}{\urlprefix }}%
\providecommand \urlprefix  [0]{URL }%
\providecommand \Eprint [0]{\href }%
\providecommand \doibase [0]{http://dx.doi.org/}%
\providecommand \selectlanguage [0]{\@gobble}%
\providecommand \bibinfo  [0]{\@secondoftwo}%
\providecommand \bibfield  [0]{\@secondoftwo}%
\providecommand \translation [1]{[#1]}%
\providecommand \BibitemOpen [0]{}%
\providecommand \bibitemStop [0]{}%
\providecommand \bibitemNoStop [0]{.\EOS\space}%
\providecommand \EOS [0]{\spacefactor3000\relax}%
\providecommand \BibitemShut  [1]{\csname bibitem#1\endcsname}%
\let\auto@bib@innerbib\@empty
\bibitem [{\citenamefont {Lamata}\ \emph {et~al.}(2007)\citenamefont {Lamata},
  \citenamefont {Le\'{o}n}, \citenamefont {Salgado},\ and\ \citenamefont
  {Solano}}]{lamata_inductive_2007}%
  \BibitemOpen
  \bibfield  {author} {\bibinfo {author} {\bibfnamefont {L.}~\bibnamefont
  {Lamata}}, \bibinfo {author} {\bibfnamefont {J.}~\bibnamefont {Le\'{o}n}},
  \bibinfo {author} {\bibfnamefont {D.}~\bibnamefont {Salgado}}, \ and\
  \bibinfo {author} {\bibfnamefont {E.}~\bibnamefont {Solano}},\ }\href
  {\doibase 10.1103/PhysRevA.75.022318} {\bibfield  {journal} {\bibinfo
  {journal} {Physical Review A}\ }\textbf {\bibinfo {volume} {75}},\ \bibinfo
  {pages} {022318} (\bibinfo {year} {2007})}\BibitemShut {NoStop}%
\bibitem [{\citenamefont {Bennett}\ \emph {et~al.}(2000)\citenamefont
  {Bennett}, \citenamefont {Popescu}, \citenamefont {Rohrlich}, \citenamefont
  {Smolin},\ and\ \citenamefont {Thapliyal}}]{bennett_exact_2000}%
  \BibitemOpen
  \bibfield  {author} {\bibinfo {author} {\bibfnamefont {C.~H.}\ \bibnamefont
  {Bennett}}, \bibinfo {author} {\bibfnamefont {S.}~\bibnamefont {Popescu}},
  \bibinfo {author} {\bibfnamefont {D.}~\bibnamefont {Rohrlich}}, \bibinfo
  {author} {\bibfnamefont {J.~A.}\ \bibnamefont {Smolin}}, \ and\ \bibinfo
  {author} {\bibfnamefont {A.~V.}\ \bibnamefont {Thapliyal}},\ }\href {\doibase
  10.1103/PhysRevA.63.012307} {\bibfield  {journal} {\bibinfo  {journal}
  {Physical Review A}\ }\textbf {\bibinfo {volume} {63}},\ \bibinfo {pages}
  {012307} (\bibinfo {year} {2000})}\BibitemShut {NoStop}%
\bibitem [{\citenamefont {D\"{u}r}\ \emph {et~al.}(2000)\citenamefont
  {D\"{u}r}, \citenamefont {Vidal},\ and\ \citenamefont
  {Cirac}}]{dur_three_2000}%
  \BibitemOpen
  \bibfield  {author} {\bibinfo {author} {\bibfnamefont {W.}~\bibnamefont
  {D\"{u}r}}, \bibinfo {author} {\bibfnamefont {G.}~\bibnamefont {Vidal}}, \
  and\ \bibinfo {author} {\bibfnamefont {J.~I.}\ \bibnamefont {Cirac}},\ }\href
  {\doibase 10.1103/PhysRevA.62.062314} {\bibfield  {journal} {\bibinfo
  {journal} {Physical Review A}\ }\textbf {\bibinfo {volume} {62}},\ \bibinfo
  {pages} {062314} (\bibinfo {year} {2000})}\BibitemShut {NoStop}%
\bibitem [{\citenamefont {Verstraete}\ \emph {et~al.}(2002)\citenamefont
  {Verstraete}, \citenamefont {Dehaene}, \citenamefont {De~Moor},\ and\
  \citenamefont {Verschelde}}]{verstraete_four_2002}%
  \BibitemOpen
  \bibfield  {author} {\bibinfo {author} {\bibfnamefont {F.}~\bibnamefont
  {Verstraete}}, \bibinfo {author} {\bibfnamefont {J.}~\bibnamefont {Dehaene}},
  \bibinfo {author} {\bibfnamefont {B.}~\bibnamefont {De~Moor}}, \ and\
  \bibinfo {author} {\bibfnamefont {H.}~\bibnamefont {Verschelde}},\ }\href
  {\doibase 10.1103/PhysRevA.65.052112} {\bibfield  {journal} {\bibinfo
  {journal} {Physical Review A}\ }\textbf {\bibinfo {volume} {65}},\ \bibinfo
  {pages} {052112} (\bibinfo {year} {2002})}\BibitemShut {NoStop}%
\bibitem [{\citenamefont {Lamata}\ \emph {et~al.}(2006)\citenamefont {Lamata},
  \citenamefont {Le\'{o}n}, \citenamefont {Salgado},\ and\ \citenamefont
  {Solano}}]{lamata_inductive_2006}%
  \BibitemOpen
  \bibfield  {author} {\bibinfo {author} {\bibfnamefont {L.}~\bibnamefont
  {Lamata}}, \bibinfo {author} {\bibfnamefont {J.}~\bibnamefont {Le\'{o}n}},
  \bibinfo {author} {\bibfnamefont {D.}~\bibnamefont {Salgado}}, \ and\
  \bibinfo {author} {\bibfnamefont {E.}~\bibnamefont {Solano}},\ }\href
  {\doibase 10.1103/PhysRevA.74.052336} {\bibfield  {journal} {\bibinfo
  {journal} {Physical Review A}\ }\textbf {\bibinfo {volume} {74}},\ \bibinfo
  {pages} {052336} (\bibinfo {year} {2006})}\BibitemShut {NoStop}%
\bibitem [{\citenamefont {Li}\ \emph {et~al.}(2006)\citenamefont {Li},
  \citenamefont {Li}, \citenamefont {Huang},\ and\ \citenamefont
  {Li}}]{li_simple_2006}%
  \BibitemOpen
  \bibfield  {author} {\bibinfo {author} {\bibfnamefont {D.}~\bibnamefont
  {Li}}, \bibinfo {author} {\bibfnamefont {X.}~\bibnamefont {Li}}, \bibinfo
  {author} {\bibfnamefont {H.}~\bibnamefont {Huang}}, \ and\ \bibinfo {author}
  {\bibfnamefont {X.}~\bibnamefont {Li}},\ }\href {\doibase
  10.1016/j.physleta.2006.07.004} {\bibfield  {journal} {\bibinfo  {journal}
  {Physics Letters A}\ }\textbf {\bibinfo {volume} {359}},\ \bibinfo {pages}
  {428} (\bibinfo {year} {2006})}\BibitemShut {NoStop}%
\bibitem [{Note1()}]{Note1}%
  \BibitemOpen
  \bibinfo {note} {Lamata et al.\ pick $\protect \mathfrak {W}_{GHZ,W}$ as the
  default class instead of $\protect \mathfrak {W}_{GHZ, GHZ}$ even though
  usually GHZ is ranked before W.}\BibitemShut {Stop}%
\bibitem [{\citenamefont {Stein~et al.}(2016)}]{stein_sage_2016}%
  \BibitemOpen
  \bibfield  {author} {\bibinfo {author} {\bibfnamefont {W.~A.}\ \bibnamefont
  {Stein~et al.}},\ }\href@noop {} {\enquote {\bibinfo {title} {{Sage
  Mathematics Software (Version 7.2)}},}\ }\bibinfo {howpublished} {The Sage
  Development Team, \url{http://www.sagemath.org}} (\bibinfo {year}
  {2016})\BibitemShut {NoStop}%
\end{thebibliography}%

\end{document}